\documentclass[preprintnumbers,twocolumn,superscriptaddress,aps,nofootinbib]{revtex4-1}
\usepackage{amssymb}
\usepackage[centertags]{amsmath}
\usepackage{txfonts}
\usepackage{epsfig}
\usepackage{bm}
\usepackage{color}
\usepackage{graphicx,graphics}
\usepackage{float}
\usepackage[pdfstartview=FitH]{hyperref}
\hypersetup{colorlinks=true, citecolor=blue, linkcolor=blue,filecolor=black,urlcolor=blue}
\allowdisplaybreaks[2]



\begin{document}

\title{New insights into hadron production mechanism from $p_{T}$ spectra in $pp$ collisions at $\sqrt{s}=7$ TeV}

\author{Xing-rui Gou}

\affiliation{Department of Physics, Qufu Normal University, Shandong 273165, China}

\author{Feng-lan Shao}
\email{shaofl@mail.sdu.edu.cn}

\affiliation{Department of Physics, Qufu Normal University, Shandong 273165, China}

\author{Rui-qin Wang}

\affiliation{Department of Physics, Qufu Normal University, Shandong 273165, China}

\author{Hai-hong Li}

\affiliation{Department of Physics, Jining University, Shandong 273155, China}

\author{Jun Song }
\email{songjun2011@jnxy.edu.cn}

\affiliation{Department of Physics, Jining University, Shandong 273155, China}

\begin{abstract}
We show that the experimental data of mid-rapidity $p_{T}$ spectra for proton, $\Lambda$, $\Xi$ , $\Omega^{-}$, $\text{K(892)}^{*0}$ and $\Xi(1530)^{*0}$ in minimum-bias $pp$ collisions at $\sqrt{s}=7$ TeV can be systemically explained by the quark combination mechanism of hadronization. The averaged transverse momentum $\langle p_{T}\rangle$ and spectra ratios such as $\Xi/\Lambda$ and $\Omega/\phi$ calculated from quark combination reproduce the data much better than those from traditional string and/or cluster fragmentation. The available data of hadronic $p_{T}$ spectra released by ALICE collaboration in the first three high-multiplicity classes of $pp$ collisions at $\sqrt{s}=7$ TeV are also well explained. We make predictions for other hadrons, and propose two scaling behaviors among decuplet baryons and vector mesons as the effective probe of hadron production mechanism at such high collision energy. 
\end{abstract}

\maketitle

\section{Introduction}

At sufficiently high temperature and energy density, nuclear matter undergoes a transition to a phase in which quarks and gluons are not confined: the quark\textendash gluon plasma (QGP) \cite{Shuryak:1980tp}.  This deconfined state is usually believed to be formed (with volume of thousand cubic fermi) in ultra-relativistic heavy ion collisions.  Unexpectedly, recent ALICE and CMS experiments at Large Hadron Collider (LHC) have revealed a series of interesting properties of hadron production in high multiplicity events of $pp$ and $p$-Pb collisions, e.~g., long range angular correlations \cite{Khachatryan:2010gv,CMS:2012qk} and collectivity \cite{Khachatryan:2015waa,Khachatryan:2016txc,Ortiz:2014iva}, enhanced strangeness \cite{Adam:2015vsf,ALICE:2017jyt} and enhanced baryon to meson ratios at soft transverse momentum \cite{Abelev:2013haa,Adam:2016dau}, which in heavy ion collisions are typically attributed to the formation of a strongly interacting QGP. Remarkable similarities in $pp$, $p$-Pb and Pb-Pb collisions at LHC have invoked intensive discussions in literature involving the mini-QGP or phase transition \cite{Liu:2011dk,Werner:2010ss,Bzdak:2013zma,Bozek:2013uha,Prasad:2009bx,Avsar:2010rf}, multiple parton interaction \cite{Sjostrand:1987su}, string overlap and color re-connection at hadronization \cite{Bautista:2015kwa,Bierlich:2014xba,Ortiz:2013yxa,Christiansen:2015yqa}, etc., in the small system created in $pp$ and $p$-Pb collisions.  The search of other new features of hadron production is important to gain deep insights into the property and hadronization of the partonic systems created in $pp$ and $p$-Pb collisions at LHC.

A series of measurements of transverse momentum $p_{T}$ spectra of identified hadrons have been carried out and high-precision data have been released by ALICE and CMS collaborations \cite{Abelev:2013haa,Adam:2016dau,Adam:2015qaa,Abelev:2012jp,Abelev:2012hy,Abelev:2014qqa,Adam:2016bpr,Adamova:2017elh}.  It is of particular interest to see that these data show any regularities that may lead to deeper insights. In the latest work \cite{Song:2017gcz}, we found that the data of $p_{T}$ spectra of identified hadrons in $p$-Pb collisions at $\sqrt{s_{NN}}=5.02$ TeV \cite{Adam:2015vsf,Abelev:2013haa,Adam:2016dau,Adam:2016bpr,Adamova:2017elh} released by ALICE collaboration exhibit a striking quark number scaling.  This scaling property is a direct consequence of quark (re-)combination mechanism (QCM) at hadronization \cite{Bjorken:1973mh,Das:1977cp,Greco:2003xt,Fries:2003vb,Hwa:2002tu,Shao:2004cn}, and it indicates the constituent quark degrees of freedom play an important rule in hadron production of small systems produced at LHC energies. We surprisingly found that quark number scaling seems to also hold in 40-60\% and 60-80\% multiplicity classes of $p$-Pb collisions where the charged-particle multiplicity density at mid-rapidity is relatively small, i.e.,$\langle dN_{ch}/d\eta\rangle\sim10-20$. Considering that $\langle dN_{ch}/d\eta\rangle$ in high-multiplicity events of $pp$ collisions at $\sqrt{s}=7$ TeV also reach such values and the hadron production in there also exhibits remarkable similarities with those in $p$-Pb collisions at $\sqrt{s_{NN}}=5.02$ TeV \cite{Khachatryan:2010gv,CMS:2012qk,Khachatryan:2015waa,Khachatryan:2016txc,Ortiz:2014iva}, we therefore further study in this paper whether quark combination also works in $pp$ collisions at $\sqrt{s}=7$ TeV or not. 

The paper is organized as follows: Sec.~II will gives a short introduction of quark (re-)combination mechanism for hadronization. Sec.~III and Sec.~IV present our results and relevant discussions in minimum-bias events and high-multiplicity events, respectively. Sec. V present the discussions on the quark number scaling property for $p_T$ spectra of vector mesons and decuplet baryons. Summary is given at last in Sec. VI. 

\section{hadron yields and $p_{T}$ spectra in QCM }

The application of QCM to the production of hadrons in high energy reactions has a long history \cite{Bjorken:1973mh,Das:1977cp}. QCM describes the formation of hadrons at hadronization by the combination of quarks and antiquarks neighboring in phase space. The mechanism assumes the effective absence of soft gluon quanta at hadronization and the effective degrees of freedom of QCD matter are only quarks and antiquarks. A quark and an antiquark neighboring in phase space form a meson and three quarks (antiquarks) form a baryon (antibaryon). Relativistic heavy-ion collisions produce a large volume of deconfined quark matter and this makes QCM to be a natural scenario for hadronization \cite{Greco:2003xt,Fries:2003vb,Hwa:2002tu,Shao:2004cn,Zimanyi:1999py,Chen:2006vc,Shao:2009uk,Song:2013isa,Wang:2012cw,Wang:2013duu,Wang:2013pnn}.

The high multiplicity events of $pp$ and $p$-Pb collisions at LHC show remarkable similarities with those of Pb-Pb collisions \cite{Khachatryan:2010gv,CMS:2012qk,Ortiz:2014iva,Adam:2015vsf,ALICE:2017jyt}.  The origin of such similarity is possibly attributed to the formation of dense parton system \cite{Liu:2011dk,Werner:2010ss,Bzdak:2013zma,Bozek:2013uha,Prasad:2009bx,Avsar:2010rf} in terms of string overlap or percolation and MPI \cite{Sjostrand:1987su,Bautista:2015kwa,Bierlich:2014xba,Ortiz:2013yxa,Christiansen:2015yqa}, etc. If such dense system is in a QGP-like deconfined state, we prefer to apply QCM to explain the data of $pp$ and $p$-Pb collisions.  In addition, the effects of hadronic re-scatterings are expected to be small for such small systems, and therefore we can get more direct information on the property of the created partonic system and its hadronization. 

When apply QCM to the small parton system, in principle, we should
follow e.g. procedures in Refs. \cite{Xie:1988wi,Wang:1996jy} by starting
from the partons after perturbative evolution to study how to treat
them or evolve them in the subsequent non-perturbative stage as a
collection of constituent quarks and antiquarks, and finally study
how to recombine them into different identified hadrons. However,
the current understanding on the multi-parton system produced at such
high collisions energies is still incomplete, in particular, for high-multiplicity
events. Therefore, in this paper we only test the basic characteristics
of QCM in $pp$ collisions at $\sqrt{s}=7$ TeV, that is, we formulate
the $p_{T}$ spectra of hadrons based on a quark statistic method
with the effective constituent quark degrees of freedom. 

As formulated in e.g. \cite{Wang:2012cw}, in general, in QCM, for a baryon $B_{j}$ composed of $q_{1}q_{2}q_{3}$ and a meson $M_{j}$
composed of $q_{1}\bar{q}_{2}$, we have 
\begin{align}
f_{B_{j}}(p_{B}) & =\int dp_{1}dp_{2}dp_{3}{\cal R}_{B_{j}}(p_{1},p_{2},p_{3};p_{B})\,f_{q_{1}q_{2}q_{3}}(p_{1},p_{2},p_{3}),\label{eq:fb}\\
f_{M_{j}}(p_{M}) & =\int dp_{1}dp_{2}{\cal R}_{M_{j}}(p_{1},p_{2};p_{M})f_{q_{1}\bar{q}_{2}}(p_{1},p_{2}),\label{eq:fm}
\end{align}
where $f_{q_{1}q_{2}q_{3}}(p_{1},p_{2},p_{3})$ is the joint momentum distribution for $q_{1}$, $q_{2}$ and $q_{3}$; and ${\cal R}_{B_{j}}(p_{1},p_{2},p_{3};p_{B})$ is the combination function that is the probability for a given $q_{1}q_{2}q_{3}$ with momenta $p_{1},p_{2}$ and $p_{3}$ to combine into a baryon $B_{j}$ with momentum $p_{B}$; and similar for mesons. If we assume independent distributions of quarks and/or antiquarks, we have 
\begin{align}
f_{q_{1}q_{2}q_{3}}(p_{1},p_{2},p_{3}) & =N_{q_{1}q_{2}q_{3}}f_{q_{1}}^{\left(n\right)}(p_{1})f_{q_{2}}^{\left(n\right)}(p_{2})f_{q_{3}}^{\left(n\right)}(p_{3}),\label{eq:fqqq}\\
f_{q_{1}\bar{q}_{2}}(p_{1},p_{2}) & =N_{q_{1}\bar{q}_{2}}f_{q_{1}}^{\left(n\right)}(p_{1})f_{\bar{q}_{2}}^{\left(n\right)}(p_{2}).\label{eq:fqqbar}
\end{align}
Here $f_{q}^{\left(n\right)}\left(p\right)$ is the single quark distribution with normalization $\int dpf_{q}^{\left(n\right)}\left(p\right)=1$ and the number of quarks of flavor $q_{i}$ is denoted by $N_{q_{i}}$.  $N_{q_{1}\bar{q}_{2}}=N_{q_{1}}N_{\bar{q}_{2}}$ is the number of possible $q_{1}\bar{q}_{2}$ pairs. $N_{q_{1}q_{2}q_{3}}=\int dp_{1}dp_{2}dp_{3}\,f_{q_{1}q_{2}q_{3}}(p_{1},p_{2},p_{3})$ is the number of three quark combinations and takes to be $6N_{q_{1}}N_{q_{2}}N_{q_{3}}$, $3N_{q_{1}}\left(N_{q_{1}}-1\right)N_{q_{2}}$ and $N_{q_{1}}\left(N_{q_{1}}-1\right)\left(N_{q_{1}}-2\right)$ for cases of three different flavors, two identical flavors and three identical flavors, respectively. Factors 6 and 3 are numbers of permutations for $q_{1}q_{2}q_{3}$ and $q_{1}q_{1}q_{2}$ combinations, respectively.  We emphasize that the form of $N_{q_{1}q_{2}q_{3}}$ has consider some necessary threshold effects for identified hadrons. For example, in $\Omega^{-}$formation $N_{sss}=N_{s}\left(N_{s}-1\right)\left(N_{s}-2\right)$ means that $\Omega^{-}$ can be only produced in events with strange quark number $N_{s}\geq3$. 

Suppose the combination takes place mainly for quark and/or antiquark that takes a given fraction of momentum of the hadron, we write the
combination function
\begin{align}
{\cal R}_{B_{j}}(p_{1},p_{2},p_{3};p_{B}) & =\kappa_{B_{j}}\prod_{i=1}^{3}\delta(p_{i}-x_{i}p_{B}),\label{eq:RB}\\
{\cal R}_{M_{j}}(p_{1},p_{2};p_{M}) & =\kappa_{M_{j}}\prod_{i=1}^{2}\delta(p_{i}-x_{i}p_{M}).\label{eq:RM}
\end{align}
 Inspired by the latest work in $p$-Pb collisions at LHC \cite{Song:2017gcz}, we adopt the approximation of equal transverse velocity in combination, or called co-moving approximation, since we apply the concept of constituent quark structure of hadrons. We recall the velocity is $v=p/E=p/\gamma m$.  Equal velocity implies $p_{i}=\gamma vm_{i}\propto m_{i}$ that leads to 
\begin{equation}
x_{i}=m_{i}/\sum_{j}m_{j},
\end{equation}
where quark masses are taken to be $m_{s}=500$ MeV and $m_{u}=m_{d}=330$ MeV.

Inserting Eqs. (\ref{eq:fqqq}-\ref{eq:fqqbar}) and (\ref{eq:RB}-\ref{eq:RM}), we obtain 
\begin{align}
f_{B_{j}}(p_{B}) & =N_{q_{1}q_{2}q_{3}}\kappa_{B_{j}}f_{q_{1}}^{\left(n\right)}(x_{1}p_{B})f_{q_{2}}^{\left(n\right)}(x_{2}p_{B})f_{q_{3}}^{\left(n\right)}(x_{3}p_{B}),\label{eq:fBz}\\
f_{M_{j}}(p_{M}) & =N_{q_{1}\bar{q}_{2}}\kappa_{M_{j}}f_{q_{1}}(x_{1}p_{M})f_{\bar{q}_{2}}(x_{2}p_{M}).\label{eq:fmz}
\end{align}
By defining the normalized hadron distributions, for $B_{j}$$\left(q_{1}q_{2}q_{3}\right)$
\begin{equation}
f_{B_{j}}^{\left(n\right)}\left(p_{B}\right)=A_{B_{j}}\,f_{q_{1}}^{\left(n\right)}\left(x_{1}p_{B}\right)f_{q_{2}}^{\left(n\right)}\left(x_{2}p_{B}\right)f_{q_{3}}^{\left(n\right)}\left(x_{3}p_{B}\right),\label{eq:fnbi}
\end{equation}
and for $M_{j}\left(q_{1}\bar{q}_{2}\right)$ 
\begin{equation}
f_{M_{j}}^{\left(n\right)}\left(p_{M}\right)=A_{M_{j}}f_{q_{1}}^{\left(n\right)}\left(x_{1}p_{M}\right)f_{\bar{q}_{2}}^{\left(n\right)}\left(x_{2}p_{M}\right),\label{eq:fnmi}
\end{equation}
where $A_{B_{j}}^{-1}=\int{\rm d}p\prod_{i=1}^{3}f_{q_{i}}^{\left(n\right)}\left(x_{i}p\right)$
and $A_{M_{j}}^{-1}=\int{\rm d}pf_{q_{1}}^{\left(n\right)}\left(x_{1}p\right)f_{\bar{q}_{2}}^{\left(n\right)}\left(x_{2}p\right)$,
we finally obtain the following formula of hadronic spectra 
\begin{align}
f_{B_{j}}\left(p_{B}\right) & =N_{B_{j}}\,f_{B_{j}}^{\left(n\right)}\left(p_{B}\right),\label{eq:fbfinal}\\
f_{M_{j}}\left(p_{M}\right) & =N_{M_{j}}\,f_{M_{j}}^{\left(n\right)}\left(p_{M}\right),\label{eq:fmfinal}
\end{align}
where yields of baryon and meson 
\begin{align}
N_{B_{j}} & =N_{q_{1}q_{2}q_{3}}P_{q_{1}q_{2}q_{3}\rightarrow B_{j}}=N_{q_{1}q_{2}q_{3}}\frac{\kappa_{B_{j}}}{A_{B_{j}}},\\
N_{M_{j}} & =N_{q_{1}\bar{q}_{2}}P_{q_{1}\bar{q}_{2}\rightarrow M_{j}}=N_{q_{1}\bar{q}_{2}}\frac{\kappa_{M_{j}}}{A_{M_{j}}}.
\end{align}
We see that $\kappa_{B_{j}}/A_{B_{j}}$ is nothing but the momentum-integrated probability of $q_{1}q_{2}q_{3}\rightarrow B_{j}$ and $\kappa_{M_{j}}/A_{M_{j}}$ is the probability of $q_{1}\bar{q_{2}}\rightarrow M_{j}$. If we take the approximation that the probability for $q\bar{q}$ to form a meson and a $qqq$ to form a baryon is flavor independent, the combination probability can be determined with a few parameters

\begin{eqnarray}
P_{q_{1}q_{2}q_{3}\rightarrow B_{j}} & = & C_{B_{j}}\frac{\overline{N}_{B}}{N_{qqq}},\label{prob_B}\\
P_{q_{1}\bar{q}_{2}\rightarrow M_{j}} & = & C_{M_{j}}\frac{\overline{N}_{M}}{N_{q\bar{q}}},\label{prob_M}
\end{eqnarray}
where $\overline{N}_{B}/N_{qqq}$ denotes the average probability of three quarks combining into a baryon and $C_{B_{j}}$ is the branch ratio to $B_{j}$ for a given flavor $q_{1}q_{2}q_{3}$ combination.  $\overline{N}_{B}=\sum_{j}\overline{N}_{B_{j}}$ is the average number of total baryons and $N_{qqq}=N_{q}(N_{q}-1)(N_{q}-2)$ is the total possible number of three quark combinations for baryon formation.  $N_{q}=\sum_{q_i}N_{q_i}$ is the total quark number. Similarly, $\overline{N}_{M}/N_{q\bar{q}}$ is used to approximately denote the average probability of a quark and antiquark combining into a meson and $C_{M_{j}}$ is the branch ratio to $M_{j}$ for a given flavor $q_{1}\bar{q}_{2}$ combination.  $\overline{N}_{M}=\sum_{j}\overline{N}_{M_{j}}$ is total mesons and $N_{q\bar{q}}=N_{q}N_{\bar{q}}$ is the total possible number of quark antiquark pairs for meson formation. 

Here we consider only the ground state $J^{P}=0^{-},\,1^{-}$ mesons and $J^{P}=(1/2)^{+},\,(3/2)^{+}$ baryons in flavor SU(3) group.
For mesons 
\begin{equation}
C_{M_{j}}=\left\{ \begin{array}{ll}
\frac{1}{1+R_{V/P}} & \text{for }J^{P}=0^{-}\textrm{ mesons}\\
\frac{R_{V/P}}{1+R_{V/P}} & \textrm{for }J^{P}=1^{-}\textrm{ mesons},
\end{array}\right.
\end{equation}
where the parameter $R_{V/P}$ represents the ratio of the $J^{P}=1^{-}$ vector mesons to the $J^{P}=0^{-}$ pseudoscalar mesons of the same flavor composition; for baryons 
\begin{equation}
C_{B_{j}}=\left\{ \begin{array}{ll}
\frac{R_{O/D}}{1+R_{O/D}} & \textrm{for }J^{P}=({1}/{2})^{+}\textrm{ baryons}\\
\frac{1}{1+R_{O/D}} & \textrm{for }J^{P}=({3}/{2})^{+}\textrm{ baryons},
\end{array}\right.
\end{equation}
except that $C_{\Lambda}=C_{\Sigma^{0}}={R_{O/D}}/{(1+2R_{O/D})},~C_{\Sigma^{*0}}={1}/{(1+2R_{O/D})},~C_{\Delta^{++}}=C_{\Delta^{-}}=C_{\Omega^{-}}=1$.
Here, $R_{O/D}$ stands for the ratio of the $J^{P}=(1/2)^{+}$ octet to the $J^{P}=(3/2)^{+}$ decuplet baryons of the same flavor composition.  Here, $R_{V/P}$ and $R_{O/D}$ are set to be 0.45 and 2.5, respectively, which are slightly different from Ref. \cite{Shao:2017eok}, in order to better tune the yields of vector mesons and decuplet baryons. The fraction of baryons relative to mesons is $N_{B}/N_{M}\approx0.085$ \cite{Song:2013isa,Shao:2017eok}. Using the unitarity constraint of hadronization $N_{M}+3N_{B}=N_{q}$, $N_{B_{j}}$ and $N_{M_{j}}$ cab be calculated at given quark numbers at hadronization. 

\section{Results in minimum-bias $pp$ collisions }

The data of mid-rapidity $p_{T}$ spectra for pion, kaon, proton, $\Lambda$, $\Xi$, $\Omega$, $\phi$, $\text{K}\left(892\right)^{*0}$, $\Sigma\left(1385\right)^{*}$, $\Xi\left(1530\right)^{*}$ are all available for minimum-bias events \cite{Adam:2015qaa,Abelev:2012jp,Abelev:2012hy,Abelev:2014qqa}.  Constraining ourselves to the mid-rapidity region $y=0$, we apply the formula in previous section to the one-dimensional $p_{T}$ space and study to what extent QCM feature exhibits in these data.

\begin{figure*}[t]
\includegraphics[scale=0.8]{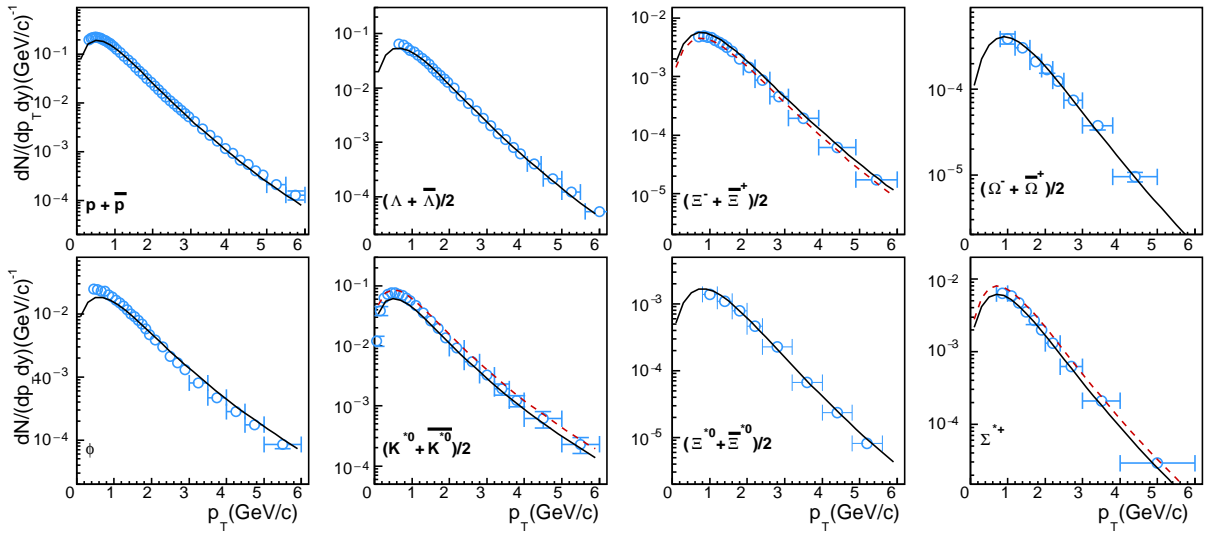}\caption{Mid-rapidity $p_{T}$ spectra of identified hadrons in minimum-bias $pp$ collisions at $\sqrt{s}=7$ TeV. The solid lines are QCM results and symbols are experimental data \cite{Adam:2015qaa,Abelev:2012jp,Abelev:2012hy,Abelev:2014qqa}.  The dashed lines are QCM results multiplied by proper constants to compare the distribution shapes. \label{fig1}}
\end{figure*}

\subsection{Considering quark number fluctuations}

To compare with the data, we should consider the large fluctuation in minimum-bias events. In particular, a rough estimation from the yield data of single-strangeness hadrons kaon and $\Lambda$ gives that the average number of strange quarks in the unit rapidity interval is only about 0.8. Since the formation of multi-strangeness hyperons such as $\Omega^{-}$ (or $\Xi$) is possible only for events with $N_{s}\geq3$ (or 2), yields of those hyperons will be strongly dependent on the fluctuation property (distribution) of strange quark number.  We get the event-averaged hadron yield by
\begin{equation}
\langle N_{h_{j}}\rangle=\sum_{\{N_{q_{i}},N_{\bar{q}_{i}}\}}\mathcal{P}\left(\left\{ N_{q_{i}},N_{\bar{q}_{i}}\right\} ;\left\{ \langle N_{q_{i}}\rangle,\langle N_{\bar{q}_{i}}\rangle\right\} \right)\,N_{h_{i}},\label{nh_pdf}
\end{equation}
where $\mathcal{P}\left(\left\{ N_{q_{i}},N_{\bar{q}_{i}}\right\} ;\left\{ \langle N_{q_{i}}\rangle,\langle N_{\bar{q}_{i}}\rangle\right\} \right)$
is the distribution of quark numbers and antiquark numbers. In this paper, we suppose the independent distribution for each flavor of quarks and antiquarks, i.e., $\mathcal{P}\left(\left\{ N_{q_{i}},N_{\bar{q}_{i}}\right\} ;\left\{ \langle N_{q_{i}}\rangle,\langle N_{\bar{q}_{i}}\rangle\right\} \right)=\prod_{q_i}\mathcal{P}\left(N_{q_i},\langle N_{q_i}\rangle\right)$.  For quark number distribution $\mathcal{P}\left(N_{q_i},\langle N_{q_i}\rangle\right)$ of specific flavor $q_i$, we firstly adopt the Poisson distribution $\text{Pois}(N_{q_i};\langle N_{q_i}\rangle)$ as a reference shape for quark number and then introduce a suppression parameter $\gamma_{q_i}\leq1$ for the long tail of Poisson distribution through a piece-wise function $\Theta(N_{q_i})=\left\{ \left\{ 1,N_{q_i}<3\right\} ,\left\{ \gamma_{q_i},N_{q_i}\geq3\right\} \right\} $.  The practical distribution is $\mathcal{P}(N_{q_i};\langle N_{q_i}\rangle)=\mathcal{N}\,Pois(N_{q_i};\mu)\Theta(N_{q_i})$ where $\mathcal{N}$ is the normalization factor and $\mu$ is solved by the average constraint $\sum_{N_{q_i}}P(N_{q_i};\langle N_{q_i}\rangle)\,N_{q_i}=\langle N_{q_i}\rangle$ for given $\gamma_{q_i}$. We take $\gamma_{s}=0.6$ for strange quark (antiquark) to better tune the yields of multi-strangeness hyperons and take $\gamma_{u}=\gamma_{d}=1$ for up and down quarks. The event-by-event fluctuation for $p_{T}$ distributions of quarks and antiquarks is neglected in this paper and we just use a event-averaged $p_{T}$ distribution function for quarks and antiquarks. 

\subsection{Mid-rapidity $p_{T}$ spectra of identified hadrons}

Inspired by the L\'evy-Tsallis parameterization \cite{Tsallis1988} for the data of hadronic $p_{T}$ spectra, we use the following form to parameterize the $p_{T}$ distribution for quarks 
\begin{equation}
f_{q}^{\left(n\right)}\left(p_{T}\right)=\mathcal{N}_{q}\left(p_{T}+a_{q}\right)^{b_{q}}\left(1+\frac{\sqrt{p_{T}^{2}+m_{q}^{2}}-m_{q}}{n_{q}c_{q}}\right)^{-n_{q}},
\end{equation}
where $\mathcal{N}_{q}$ is the normalization constant satisfying $\int dp_{T}f_{q}^{\left(n\right)}\left(p_{T}\right)=1$. Parameter $a_{q}$ is introduced to tune the spectrum at very small $p_{T}$ and is taken to be 0.06 GeV. Parameters $b_{q}$, $n_{q}$ and $c_{q}$ (GeV) tune the behavior of the spectrum at low and intermediate $p_{T}$, and are taken to be (0.485, 3.93,0.28) for $u$ or $d$ quarks and (0.485,0.405,0.362) for $s$ quark. The averaged quark numbers in $|y|<0.5$ interval are taken to be (2.5, 2.5, 0.8) for $u$, $d$, and $s$ quarks. Antiquark numbers and parameters for $p_{T}$ spectra are the same as the quarks. 

We calculate the $p_{T}$ spectra of proton, $\Lambda$, $\Xi^{0}$, $\Omega^{-}$, $\phi$, K(892)$^{*0}$ , $\Xi\text{(1530)}^{*0}$ and $\Sigma\text{(1385)}^{*+}$ in minimum-bias $pp$ collisions at $\sqrt{s}=7$ TeV, and compare them with the experimental data \cite{Adam:2015qaa,Abelev:2012jp,Abelev:2012hy,Abelev:2014qqa} in Fig. \ref{fig1}. The solid lines are our results and symbols are experimental data \cite{Adam:2015qaa,Abelev:2012jp,Abelev:2012hy,Abelev:2014qqa,Khachatryan:2011tm}.  The dashed lines are results multiplied by proper constants to remove the yield under(over)-estimation to better compare the shape of $p_{T}$ spectrum. Decuplet baryons $\Omega^{-}$, $\Xi^{*}$, $\Sigma^{*}$ and vector mesons $\text{K}(892)^{*0}$ and $\phi$ are less influenced by decay, and, therefore, behaviors of these hadrons are usually believed as carrying more direct information from hadronization. We see that the spectrum shapes of $\Omega^{-}$, $\Xi^{*}$, $\Sigma^{*}(1385)$ and $\text{K}(892)^{*0}$ are reproduced very well. Result of $\phi$ is somewhat flatter than the data. Taking also the decay influence into account, results of proton, $\Lambda$, $\Xi^{0}$ are in good agreement in spectrum shapes with the experimental data. Results of pion and kaon are discussed in Appendix \ref{appendixA}. 

To further quantify our results, we calculate the yield at mid rapidity $N_{h}=\int f_{h}(p_{T})dp_{T}$ and average transverse momentum $\langle p_{T}\rangle=N_{h}^{-1}\int p_{T}\:f_{h}(p_{T})dp_{T}$, and show them in Table \ref{tab1} and compare with the available experimental data. For yields, we see that on the whole our results are in good agreement with the data. In particular, the hierarchy property among yields of $p$, $\Lambda$, $\Xi^{0}$, $\Xi^{*0}$, $\Omega^{-}$ which span three orders of magnitude is well reproduced.  For hadronic $\langle p_{T}\rangle$, we get a better agreement with the data considering the statistical and systematical uncertainties.  
\begin{table}
\caption{Yield densities and average transverse momentum $\langle p_{T}\rangle$
in minimum-bias $pp$ collisions at $\sqrt{s}=7$ TeV. Experimental
data are from \cite{Adam:2015qaa,Abelev:2012jp,Abelev:2012hy,Abelev:2014qqa,Khachatryan:2011tm}.
\label{tab1}}
\begin{tabular}{ccccc}
\toprule 
    & \multicolumn{2}{c}{$\frac{dN}{dy}(\times 10^2)$} & \multicolumn{2}{c}{$\langle p_{T}\rangle$}\\ \colrule
 & data & QCM & data & QCM\\ \colrule
$\text{K}{}^{*0}$ & $9.7\pm0.04_{\:-0.9}^{\:+1.0}$ & 8.4 & $1.01\pm0.003\pm0.02$ & 1.00\\ \colrule
$\phi$ & $3.2\pm0.04_{\,-0.35}^{\,+0.4}$ & 2.9 & $1.07\pm0.005\pm0.03$ & 1.13\\ \colrule
$p$ & $12.4\pm0.9$ & 12.1 & $0.9\pm0.029$ & 0.91\\ \colrule
$\Lambda$ & $8.1\pm1.5$ & 7.7 & $1.037\pm0.005\pm0.063$ & 1.05\\ \colrule
$\Xi^{0}$ & $0.79\pm0.01_{\,-0.05}^{\,+0.07}$ & 0.91 & $1.21\pm0.01\pm0.06$ & 1.215\\ \colrule
$\Sigma^{*+}$ & $1.0\pm0.02_{\,-0.14}^{\,+0.15}$ & 0.94 & $1.16\pm0.02\pm0.07$ & 1.14\\ \colrule
$\Xi^{*0}$ & $0.256\pm0.007_{\,-0.037}^{\,+0.040}$ & 0.28 & $1.31\pm0.02\pm0.09$ & 1.26\\ \colrule
$\Omega^{-}$ & $0.0675\pm0.003_{\,-0.006}^{\,+0.008}$ & 0.075 & $1.455\pm0.03\pm0.08$ & 1.38\\ 
\botrule
\end{tabular}
\end{table}

\subsection{Discussions on $\langle p_{T}\rangle$ and particle ratios}

\begin{figure}[H]
\includegraphics[scale=0.43]{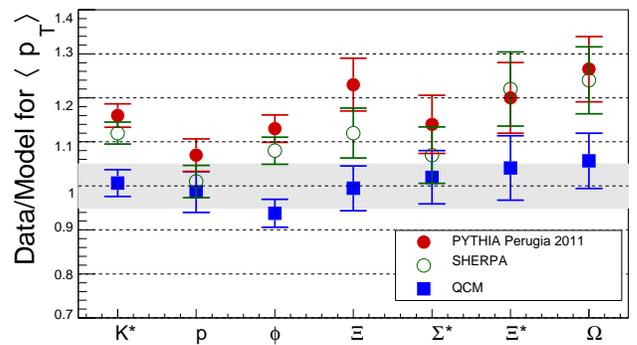}\caption{The ratios of average transverse momentum $\langle p_{T}\rangle$ for experimental data to those for models or event generators in minimum-bias $pp$ collisions at $\sqrt{s}=7$ TeV \cite{Adam:2015qaa,Abelev:2012jp,Abelev:2012hy,Abelev:2014qqa}.\label{fig2}}
\end{figure}

In Fig. \ref{fig2}, we show the ratio of data of $\langle p_{T}\rangle$ to our results for different identified hadrons, and compare them with results from different models or event generators. We see that the deviation of our results from the data is in general less than about 5\%. Popular event generator PYTHIA \cite{Sjostrand:2006za} adopts the string fragmentation \cite{Andersson:1983ia} for hadronization. Results from PYTHIA6 P2011 (tune Perugia2011) \cite{Skands:2010ak}, solid circles, show that it predicts the much soft $p_{T}$ spectra for multi-strangeness hadrons, i.e., about 20\% softer than the data. Taking effects of color re-connection into account seems to little change the $\langle p_{T}\rangle$ results \cite{Christiansen:2015yqa}. Recently, a new model of generating the transverse momentum of hadrons during the string fragmentation process, inspired by thermodynamics, can improve $\langle p_{T}\rangle$ results of multi-strangeness hyperons with degrees of about 5\% \cite{Fischer:2016zzs}. In addition, PYTHIA usually under-estimates yields of multi-strangeness hadrons $\Xi$ and $\Omega^{-}$, which however can be relieved by color re-connection \cite{Ortiz:2013yxa,Christiansen:2015yqa,Bierlich:2015rha} and/or string overlap effects realized in DIPSY \cite{Bierlich:2014xba}. Popular event generator SHERPA adopts the cluster fragmentation \cite{Webber:1983if} for hadronization.  It also predicts too soft spectra for multi-strangeness hadrons with about 15\% deviations. 

\begin{figure}[H]
    \includegraphics[scale=0.4]{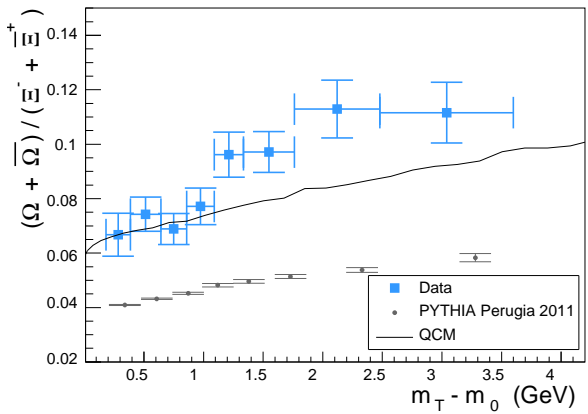}
    \caption{$\left(\Omega^{-}+\bar{\Omega}^{+}\right)$ to $\left(\Xi^{-}+\bar{\Xi}^{+}\right)$ ratio as a function of $m_{T}-m_{0}$ in minimum-bias $pp$ collisions at $\sqrt{s}=7$ TeV. Solid squares are experimental data and solid circles are PYTHIA Perugia 2011 simulation \cite{Skands:2010ak}, which are taken from \cite{Abelev:2012jp}. The line is our result. \label{fig3}}
\end{figure}

In Fig. \ref{fig3}, we show the result of $\left(\Omega^{-}+\bar{\Omega}^{+}\right)$ to $\left(\Xi^{-}+\bar{\Xi}^{+}\right)$ ratio as a function of $m_{T}-m_{0}$ in minimum-bias $pp$ collisions at $\sqrt{s}=7$ TeV, and compare it with the data \cite{Abelev:2012jp}. Our result is consistent with the data at low $m_{T}-m_{0}$ and is slightly below the data at moderate $m_{T}-m_{0}$ in magnitude. PYTHIA P2011 \cite{Skands:2010ak} predicts a too low ratio, but as mentioned above, consideration of color re-connection and string overlap will raise the ratio to a certain extent. 

\begin{figure}[H]
\includegraphics[scale=0.4]{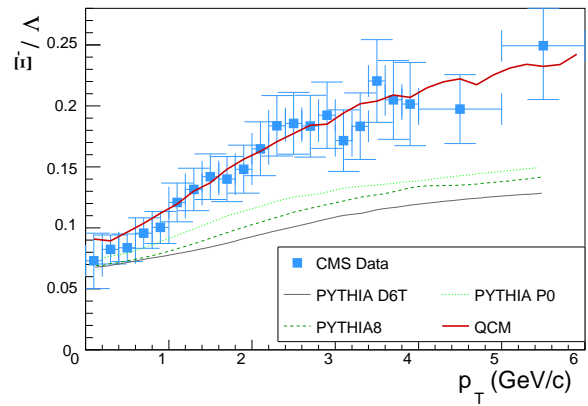}
    \caption{$\Xi^{-}/\Lambda$ ratio as a function of $p_{T}$ in minimum-bias $pp$ collisions at  $\sqrt{s}=7$ TeV. Solid squares are experimental data from CMS collaboration \cite{Khachatryan:2011tm} and thin lines are prediction of different PYTHIA versions and/or tunes \cite{Bartalini:2010su,Skands:2009zm,Sjostrand:2007gs}, which are taken from \cite{Khachatryan:2011tm}. The thick line is our result. \label{fig4b}}
\end{figure}

In Fig. \ref{fig4b} we show $\Xi^{-}/\Lambda$ ratio as a function of $p_{T}$ in minimum-bias $pp$ collisions at $\sqrt{s}=7$ TeV, and compare it with the experimental data from CMS collaboration \cite{Khachatryan:2011tm}.  Results of PYTHIA6 D6T tune \cite{Bartalini:2010su}, Perugia0 (P0) tune \cite{Skands:2009zm}, and PYTHIA8 \cite{Sjostrand:2007gs} are also shown. We see that our result is in good agreement with the data. 

\begin{figure}[H]
    \includegraphics[scale=0.4]{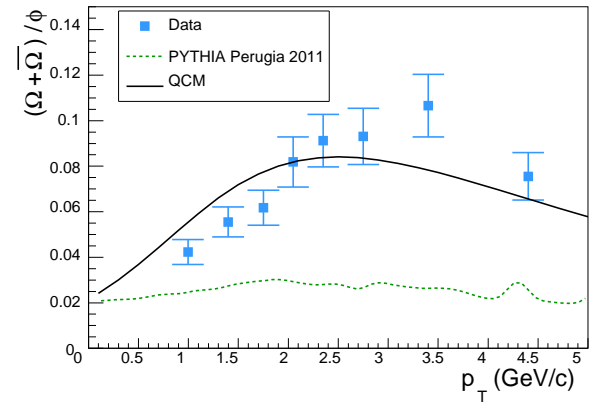}
    \caption{$\left(\Omega^{-}+\bar{\Omega}^{+}\right)$ to $\phi$ ratio as a function of $p_{T}$ in minimum-bias $pp$ collisions at $\sqrt{s}=7$ TeV. Solid squares are experimental data; dashed line is PYTHIA Perugia 2011 simulation \cite{Skands:2010ak}; they are taken from \cite{Abelev:2012hy}.  The solid line is our result. \label{fig4}}
\end{figure}

In Fig. \ref{fig4}, we show the result of $\left(\Omega^{-}+\bar{\Omega}^{+}\right)$ to $\phi$ ratio as a function of $p_{T}$ in minimum-bias $pp$ collisions at $\sqrt{s}=7$ TeV. Our result is slightly higher than the data at $p_{T}\lesssim2$ GeV and is slightly lower than the data for $p_{T}$ around 3.5 GeV, but on the whole the magnitude and shape are in good agreement with the data \cite{Abelev:2012hy}. We emphasize that such behavior of Baryon/Meson ratio is a typical property of QCM and had been observed many times in AA and pA collisions at RHIC and LHC \cite{Greco:2003xt,Fries:2003vb,Hwa:2002tu,Abelev:2006jr,Chen:2006vc}.  PYTHIA6 P2011 \cite{Skands:2010ak} predicts an obviously low and flat ratio.

\begin{figure*}
\includegraphics[scale=0.8]{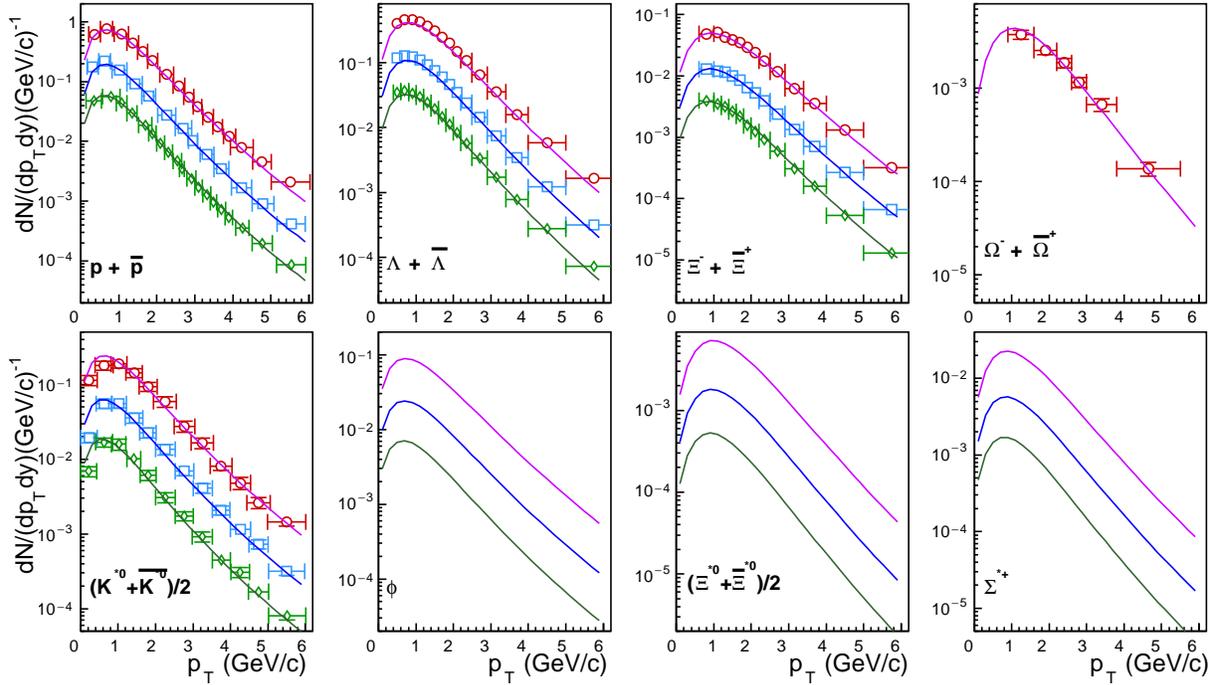}
    \caption{Mid-rapidity $p_{T}$ spectra of identified hadrons in first three high-multiplicity classes (I), (II), (III) in $pp$ collisions at $\sqrt{s}=7$ TeV. The solid lines are QCM results and symbols are experimental data. The data of $\Lambda$, $\Xi$ and $\Omega^{-}$are from \cite{ALICE:2017jyt} and those of proton and $\text{K}(892)^{*0}$ are preliminary \cite{DerradideSouza:2016kfn}. The data and our results in classes (II) and (III) are divided by factors 3 and $3^{2}$ for clarity, respectively. \label{fig5}}
\end{figure*}

In short summary of this section, on the whole we see that QCM can systematically explain the data of $p_{T}$ spectra of various identified hadrons observed in minimum-bias $pp$ collisions at $\sqrt{s}=7$ TeV. There exists small deviations for yield and/or $p_{T}$ spectra of few hadrons like $\phi$ at about 10\% level. This is might because that most events in minimum-bias event collections have small charged particle multiplicities (or quark numbers in QCM language and/or effective energy for particle production in general language). Various kinds of threshold effects may appear for these events. For example, too small $s$-quark numbers i.e. $N_{s}<3$ (or too small effective energy for $s$ quark production) in an event will inhibit the formation of $\Omega$ and/or constrain its carrying momentum. Therefore, we expect that, in high multiplicity events of $pp$ collisions where the number of quarks (or effective energy for particle production) is large and such threshold effects are weak, QCM will make better prediction for the production of identified hadrons.

\section{predictions in high-multiplicity events of $pp$ collisions}

Mid-rapidity $\langle dN_{ch}/d\eta\rangle$ in first three high-multiplicity classes 0-0.95\% (I), 0.95-4.7\% (II) and 4.7-9.5\% (III) in $pp$ collisions at LHC are $21.3\pm0.6$ , $16.5\pm0.5$ and $13.5\pm0.4$ \cite{ALICE:2017jyt}. High multiplicity means the high energy deposited in the collision region by the intense partonic interactions happening in early stage of collisions, which increases the possibility of the formation of the deconfined system. In deed, the experimental data have shown that production of hadrons in high-multiplicity events in $pp$ collisions exhibit lots of remarkable similarities with $p-$Pb collisions and Pb+Pb collisions \cite{Khachatryan:2010gv,CMS:2012qk,Ortiz:2014iva,ALICE:2017jyt}.  If QGP-like deconfined system does form in these high-multiplicity collisions, we can apply QCM in a more natural way and expect to make better predictions for the momentum spectra of identified hadrons. 

\begin{figure}
\includegraphics[scale=0.4]{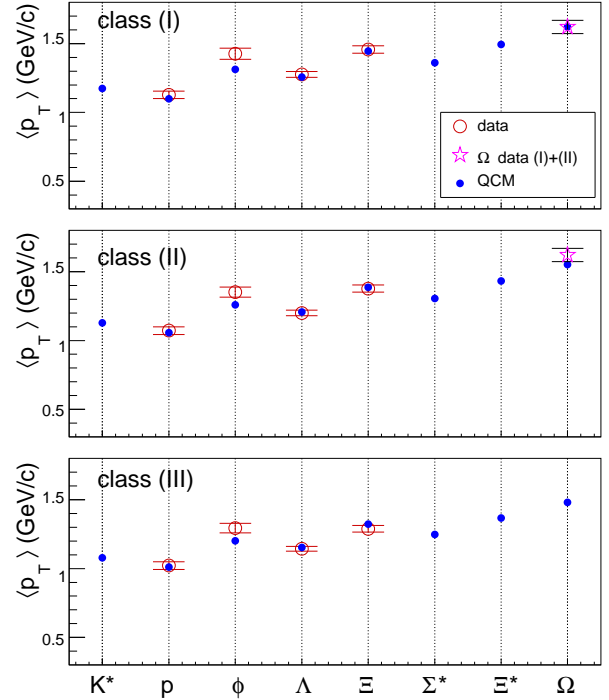}
    \caption{$\langle p_{T}\rangle$ of identified hadrons in first three high-multiplicity classes (I), (II), and (III) in $pp$ collisions at $\sqrt{s}=7$ TeV. Open symbols are experimental data \cite{ALICE:2017jyt,Jacazio:2017vcw}.\label{fig6} }
\end{figure}

We calculate the mid-rapidity $p_{T}$ spectra of identified hadrons in first three high-multiplicity classes in $pp$ collisions at $\sqrt{s}=7$ TeV, and compare them with available experimental data in Fig. \ref{fig5}.  The data of $\Lambda$, $\Xi$ and $\Omega^{-}$are from \cite{ALICE:2017jyt} and those of proton and $\text{K}(892)^{*0}$ are preliminary \cite{DerradideSouza:2016kfn}.  The fitted parameters for quark $p_{T}$ spectra (except $a_{q}=0.05$ GeV, $b_{q}=0.5$) and quark numbers are shown in Table \ref{tab3}. The average transverse momenta $\langle p_{T}\rangle$ are also calculated and are compared with available experimental data \cite{ALICE:2017jyt,Jacazio:2017vcw} in Fig. \ref{fig6}. 

From Fig. \ref{fig5}, we see that the spectrum shapes of proton, $\Lambda$, $\Xi$, $\Omega,$ and $\text{K}(892)^{*0}$ are in good agreement with the data \cite{ALICE:2017jyt,DerradideSouza:2016kfn}. Predictions of $\phi$, $\Xi^{*}$ and $\Sigma^{*+}$ are presented. Fig. \ref{fig6} shows results of $\langle p_{T}\rangle$ for proton, $\Lambda$ and $\Xi$ are in good agreement with the data. The result of $\phi$ is slightly smaller than the preliminary data \cite{Jacazio:2017vcw}. The data of $\Omega$ in multiplicity class (I)+(II) are also shown in Fig. \ref{fig6} as a visual guide. Our result of $\langle p_{T}\rangle$ for $\Omega$ in the average of classes (I)+(II) is 1.57, which is 5\% smaller than the data $1.62\pm0.05$ \cite{ALICE:2017jyt}. We note that the behaviors of $\Omega$ and $\phi$ in comparison with data in high multiplicity events are different from those in minimum-bias events shown in Table \ref{tab1}, where the $\langle p_{T}\rangle$ of $\Omega$ is slightly smaller than the data while that of $\phi$ is larger the data. 

\begin{table}
\caption{The fitted parameters $n_{q}$ and $c_{q}$ for quark $p_{T}$ spectra, quark numbers $\langle N_{u}\rangle=\langle N_{d}\rangle$ and $\langle N_{s}\rangle$ in the rapidity interval $|y|\leq0.5$ in first three multiplicity classes in $pp$ collisions at $\sqrt{s}=7$ TeV. \label{tab3}}
\begin{tabular}{ccccccc}
\toprule 
Event classes  & $n_{u}$ & $c_{u}$(GeV) & $n_{s}$ & $c_{s}$(GeV) & $\langle N_{u}\rangle$ & $\langle N_{s}\rangle$\\ \colrule
class I & 4.45  & 0.35 & 5.56 & 0.46 & 10.9 & 3.7\\ \colrule
class II & 4.25  & 0.33 & 5.26 &  0.43 & 8.4 & 2.8\\ \colrule
class III & 4.12  & 0.31 & 5.00 & 0.40 & 7.2 & 2.4\\ 
\botrule
\end{tabular}

\end{table}

\begin{figure}
\includegraphics[scale=0.4]{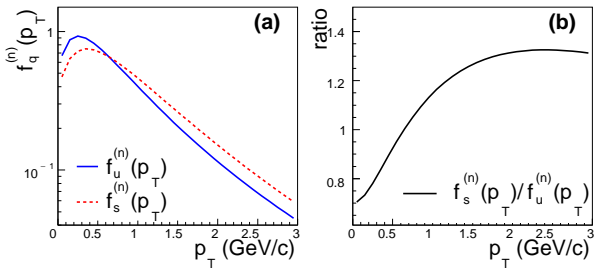}
    \caption{The $p_{T}$ spectra of $u$ and $s$ quarks in multiplicity class (I) and the ratio between them.\label{fig7}}

\end{figure}

In addition, by fitting the data of hadronic $p_{T}$ spectra, we obtain the $p_{T}$ spectra of $u$ (or $d$) quark and $s$ quark at hadronization, which show an interesting property. As an example, we plot $f_{s}^{(n)}(p_{T})$ and $f_{u}^{(n)}(p_{T})$ in the highest-multiplicity class (I) in Fig. \ref{fig7}. We see that the obtained spectrum for strange quark is harder than that for $u$ or $d$ quarks for $p_{T}$ less than 3 GeV. We also plot the ratio between them where we see it raises with $p_{T}$ and seems to reach the maximum at $p_{T}$ around 3 GeV. Results in other multiplicity classes and in minimum-bias events are similar. We note that these properties are similar to those obtained in $p$-Pb collisions at $\sqrt{s_{NN}}=5.02$ TeV \cite{Song:2017gcz} and those obtained in heavy ion collisions at RHIC and LHC energies \cite{Shao:2009uk,Wang:2013pnn,Chen:2008vr}. This is an indication of some universal property for constituent quarks evolved from the non-perturbative partonic stage. 

\begin{figure*}
\includegraphics[width=0.9\linewidth]{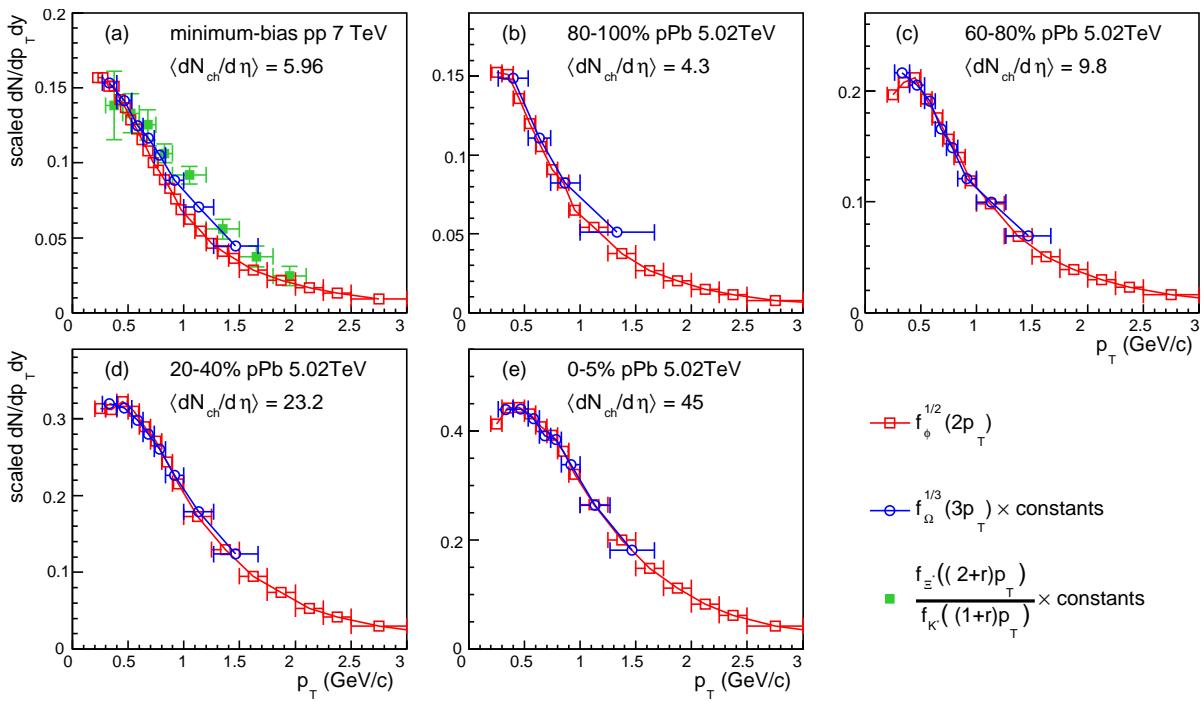}
    \caption{The scaled data for mid-rapidity $p_{T}$ spectra of $\Omega^{-}$ and $\phi$ in minimum-bias $pp$ collisions at $\sqrt{s}=7$ TeV \cite{ALICE:2017jyt,Abelev:2012hy} and those in different multiplicity classes in $p$-Pb collisions at $\sqrt{s_{NN}}=5.02$ TeV \cite{Adam:2016bpr,Adam:2015vsf}.
\label{fig9}}
\end{figure*}

\section{Discussions on quark number scaling of hadronic $p_{T}$ spectra}

If QCM indeed dominates the hadronization process of small partonic system created in $pp$ and $p$-Pb collisions, we can obtain several interesting scaling properties for $p_{T}$ spectra of identified hadron. In particular, for $\Omega^{-}$ and $\phi$ that are composed of only strange quarks (antiquarks), we have 
\begin{equation}
f_{\Omega}^{1/3}\left(3p_{T}\right)=\kappa_{\phi,\Omega}\,f_{\phi}^{1/2}\left(2p_{T}\right), \label{eq:cqsOmgPhi}
\end{equation}
where $\kappa_{\phi,\Omega}$ is a constant independent of $p_{T}$.  This is obtained directly from Eqs. (\ref{eq:fnbi}-\ref{eq:fmfinal}) and has been experimentally verified by the data of high-multiplicity $p$-Pb collisions at $\sqrt{s_{NN}}=5.02$ TeV \cite{Song:2017gcz}. However, we should emphasize that such scaling property only holds under the conditions Eqs. (\ref{eq:fqqq}) and (\ref{eq:fqqbar}) which are usually valid for a relatively large system. We also expect other scaling behaviors for decuplet baryons such as $\Xi^{*0}$ and vector mesons such as $\text{K}(892)^{*0}$, e.g., 
\begin{equation}
\frac{f_{\Xi^{*0}}\left(\left(2+r\right)p_{T}\right)}{f_{K^{*0}}\left(\left(1+r\right)p_{T}\right)}=\kappa_{\phi,K^{*},\Xi^{*}}f_{\phi}^{1/2}\left(2p_{T}\right)\label{eq:cqsXiK}
\end{equation}
 where $\kappa_{\phi,K^{*},\Xi^{*}}$ is constant. $r$ denotes the ratio of transverse momentum carried by $u$ or $d$ quark to that of $s$ quark(s), and takes to be about 2/3 in equal transverse velocity combination if we take $m_{s}=500$ MeV and $m_{u}=m_{d}=330$ MeV.  It is also experimentally verified by the data of high-multiplicity $p$-Pb collisions at $\sqrt{s_{NN}}=5.02$ TeV \cite{Song:2017gcz}.

We find that the above two scaling properties are broken for the data of minimum-bias $pp$ collisions at $\sqrt{s}=7$ TeV, as shown in Fig. \ref{fig9}(a). We see that the scaled data $f_{\Omega}^{1/3}\left(3p_{T}\right)$ is flatter than the scaled data $f_{\phi}^{1/2}\left(2p_{T}\right)$.  This, however, is not an indication for the failure of QCM but is more related to the event mix feature of minimum-bias data set. As we know, only in events with $N_{s}\geq3$ both $\Omega^{-}$ and $\phi$ can be potentially formed and in other events only $\phi$ can be formed. In general, more strange quarks produced in an event means the more intensive partonic interactions that will broaden transverse momenta, which can be inferred from the increases of $\langle p_{T}\rangle$ as the function of $\langle dN_{ch}/d\eta\rangle$ \cite{Jacazio:2017vcw}. Therefore it is reasonable to expect that the $f_{s}\left(p_{T}\right)$ in events $N_{s}\geq3$ for $\Omega^{-}$ formation is broader than that in events $N_{s}\geq1$ for $\phi$ formation, which leads the observation in Fig. \ref{fig9}(a). Because of similar reasons, the scaled data $f_{\Xi^{*}}/f_{K^{*}}$ is also flatter than the scaled data $f_{\phi}^{1/2}\left(2p_{T}\right)$.

By selecting the high-multiplicity events where the strange quark number is usually large than 3 and the above threshold effects are negligible, we can expect the restoration of quark number scaling for $p_{T}$ spectra of hadrons. A rough estimation gives that such events should have quarks and antiquarks with numbers at least $N_{s}\gtrsim3$ and $N_{u}=N_{d}\gtrsim9$ at mid-rapidities, which is corresponding to the events with the produced charged-particle density $dN_{ch}/d\eta\gtrsim20$.  Such $dN_{ch}/d\eta$ is reached in high-multiplicity class (I) in $pp$ collisions at $\sqrt{s}=7$ TeV. Unfortunately, the data of $\phi$, $\text{K}(892)^{*0}$ and $\Xi^{*}$ are unavailable at present to carry out such test. However, we have got some hint from the discussions around Fig. \ref{fig6} in the above section where we find that preliminary data for $\langle p_{T}\rangle$ of $\Omega^{-}$ and $\phi$ can be potentially reproduced by the same $s-$quark $p_{T}$ distribution.  In addition,we may get some indications also from the restoration of quark number scaling for the data of $p$-Pb collisions at $\sqrt{s_{NN}}=5.02$ TeV with the increase of the charged-particle multiplicity at mid-rapidity, which is shown in Fig. \ref{fig9}(b)-(e). 

Some comments on the relation of our works/results and the creation of the deconfined QGP-like system are necessary. Quark combination is a microscopic mechanism for hadronization and in principle it can be applied to various partonic final states created in high energy reactions. In particular, we recall that it was successfully used to explain the hadron production in $e^{+}e^{-}$, $pp$, $p\bar{p}$, and other hadron reactions in early years, see e.g. Refs \cite{Aitala:1996hf,Aitala:1997nt,Braaten:2002yt,Das:1977cp,Xie:1988wi}, where the deconfined system is not created. Therefore, the key point on the application of quark combination mechanism to study the formation of deconfined system is that whether we can find some ``free'' characteristics for the quarks and antiquarks extracted from hadron observables in specific reactions. Eqs. (\ref{eq:cqsOmgPhi}) and \ref{eq:cqsXiK} are such kinds of examples, which have been observed in high-multiplicity $p$-Pb collisions at $\sqrt{s_{NN}}=5.02$ TeV. The observation of Eqs. (\ref{eq:cqsOmgPhi}) and (\ref{eq:cqsXiK}) indicates that $s$ quark can freely combine with neighboring quarks and/or antiquarks so that it can not only form $\Omega^{-}$ but also form $\phi$, $\text{K}^{*}$, $\Xi^{*}$ etc. Together with the analysis of charged-particle multiplicity, we finally obtain that e.g, in high-multiplicity $p$-Pb collisions at $\sqrt{s_{NN}}=5.02$ TeV, there exist a underlying source with dozens of quarks and antiquarks which can freely combine with each other into hadrons at hadronization. We think it is a possible signal for the creation of deconfined system in these collisions. 

\section{Summary}

Using the quark combination mechanism for hadron formation at hadronization (QCM), we have studied the mid-rapidity $p_{T}$ spectra of identified hadrons in $pp$ collisions at $\sqrt{s}=7$ TeV. For minimum-bias events, the experimental data of $p_{T}$ spectra of identified hadrons except $\phi$ are well described. The $p_{T}$ integrated yields are in good agreement with the data. In particular the hierarchy among yields of proton, $\Lambda$, $\Xi$ and $\Omega$ with different strangeness content are well reproduced. The calculated average transverse momentum $\langle p_{T}\rangle$ deviate from the data less than 5\%, which is much smaller than results of event generators PYTHIA and SHERPA which adopt the string fragmentation and/or cluster fragmentation, respectively. The ratios of $\Omega/\Xi$, $\Xi/\Lambda$, and $\Omega/\phi$ as the function of $p_{T}$ are well described both in magnitudes and shapes. In first three classes of high-multiplicity events in $pp$ collisions where the deconfined system is most possibly created, the available data of $p_{T}$ spectra of hadrons are well reproduced by QCM. These results suggest that the constituent quark degrees of freedom play an important role for hadron production also in such small systems at LHC energy. We predict the $p_{T}$ distribution of other hadrons for the future test and propose the possible existence of two interesting scaling behaviors for the $p_{T}$ spectra of decuplet baryons and vector mesons in high-multiplicity $pp$ collisions at LHC, which have been observed in $p$-Pb collisions at $\sqrt{s_{NN}}=5.02$
TeV. 

\begin{acknowledgments}
    We thank Z. T. Liang, Z. B. Xu, Z. X. Zhang,  and J. Y. Jia for helpful discussions.  This work is supported by the National Natural Science Foundation of China under Grant Nos. 11675091, 11575100, 11505104 and 11305076.
\end{acknowledgments}

\bibliographystyle{apsrev4-1}
\bibliography{refpp}

\appendix

\section{$p_{T}$ spectra of pion and kaon\label{appendixA}}

The mass of pion is much smaller than the sum of the masses of its constituent quarks if we take $m_{u}=m_{d}=330$ MeV. There is a large energy discrepancy in direct combination $u(d)+\bar{u}(\bar{d})\rightarrow\pi$.  Similar situation occurs for pseudo-scalar kaon in $u(d)+\bar{s}\rightarrow K$ or $\bar{u}(\bar{d})+s\rightarrow\bar{K}$ process with $m_{s}=500$ MeV and the above $m_{u}$ and $m_{d}$. 

A possible phenomenological solution to this issue is that we consider 
\begin{align}
u+\bar{d} & \rightarrow R\rightarrow\pi^{+}+X,\label{eq:pio_X}\\
u+\bar{s} & \rightarrow R'\rightarrow K^{+}+X,\label{eq:kaon_X}
\end{align}
where $R$ and $R'$ are some intermediate resonances or clusters and $X$ is some soft degrees of freedom at hadronization. In this way the energy conservation is satisfied explicitly. Here $m_{R}=0.66$ GeV and $m_{R'}=0.83$ GeV are taken to be the sum of incoming constituent quarks under the equal velocity combination approximation. For $X$, if we identify them as pion we have $m_{X}=m_{\pi}$; if we identify them as soft gluons we have $m_{X}\lesssim m_{\pi}$. For combination $u+\bar{u}\rightarrow R$, we consider its two possible outgoing channels $R\rightarrow\pi^{0}+\pi^{0}$ and $R\rightarrow\pi^{+}+\pi^{-}$ with equal weight. The decays of $R$ and $R'$ are assumed to be isotropic. 

\begin{figure}
\includegraphics[scale=0.4]{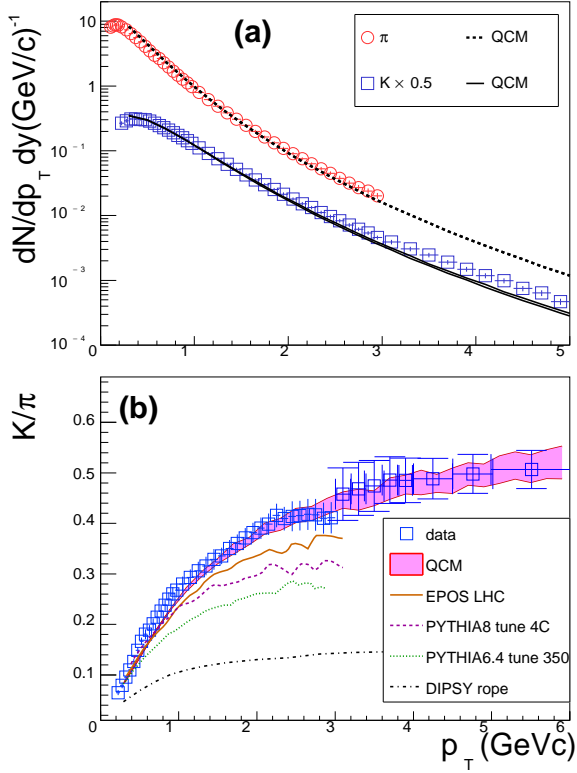}
    \caption{Mid-rapidity $p_{T}$ spectra of kaon and pion in minimum-bias $pp$ collisions at $\sqrt{s}=7$ TeV and the ratio between them. Symbols are experimental data \cite{Adam:2015qaa,Jacazio:2017vcw} and results of other models and/or event generators are taken from \cite{Bierlich:2015rha,Adam:2015qaa}.  \label{figKpi}}
\end{figure}

In Fig. \ref{figKpi}(a), we show results of $p_{T}$ spectra of kaon and pion obtained this way in minimum-bias $pp$ collisions at $\sqrt{s}=7$ TeV, in which the decay contribution from other hadrons is also included.  We find that the result of pion changes little for $0< m_{X}\leq m_{\pi}$, which is mainly because most of pions observed in experiments are from the decay of other hadrons. Our result of pion is found to be in good agreement with the available experimental data. The result of kaon changes weakly for $0< m_{X}\leq m_{\pi}$, which is shown as a very thin band in the calculated $p_{T}$ spectrum for kaon in Fig. \ref{figKpi}(a). We see that the result is in good agreement with the data of kaon for $p_{T}\lesssim2$ GeV but is lower than the data to a certain extent at moderate $p_{T}$, which may show some limitation of such a crude treatment for kaon formation in Eq.  (\ref{eq:kaon_X}) and perhaps for pion in Eq. (\ref{eq:pio_X}) also. 

In Fig. \ref{figKpi}(b), we show the ratio of kaon to pion as the function of $p_{T}$ which may cancel such limitation to a certain extent. The ratio is dependent on $m_{X}$ to a certain extent as $p_{T}\gtrsim1.5$ GeV and therefore is shown as a band corresponding to $0<m_{X}\leq m_{\pi}$. We find that it agrees well with the experimental data \cite{Adam:2015qaa,Jacazio:2017vcw}. We also show results of PYTHIA with different versions and/or tunes \cite{Sjostrand:2007gs,Skands:2010ak}, DIPSY with string overlap effect by color rope \cite{Bierlich:2015rha}, and EPOS for LHC \cite{Pierog:2013ria} as the characteristics of string fragmentation. Our results for $p_{T}$ spectra of kaon and pion in high-multiplicity classes (I), (II) and (III) are similar to those in minimum-bias events, and in particular, the $\text{K}/\pi$ ratios are also well explained.

\end{document}